# An ubiquitous ~62 Myr periodic fluctuation superimposed on general trends in fossil biodiversity

A summary of comments at the North American Paleontological Convention, June 25 2009.

*Adrian L. Melott, Department of Physics and Astronomy, University of Kansas, Lawrence, Kansas 66045  E-mail: melott@ku.edu*

*Richard K. Bambach, Department of Paleobiology, National Museum of Natural History, Smithsonian Institution, PO Box 37012, MRC 121, Washington, DC 20013-7012  E-mail: richard.bambach@verizon.net*

A 62 Myr periodicity is superimposed on other longer-term trends in fossil biodiversity. This cycle can be discerned in marine data based on the Sepkoski compendium, the Paleobiology Database, and the Fossil Record 2. The signal also exists in changes in sea level/sediment, but is much weaker than in biodiversity itself. A significant excess of 19 previously identified Phanerozoic mass extinctions occur on the declining phase of the 62 Myr cycle. Given the appearance of the signal in sampling-standardized biodiversity data, it is likely not to be a sampling artifact, but either a consequence of sea-level changes or an additional effect of some common cause for them both. In either case, it is intriguing why both changes would have a regular pattern.

Part I:
We use Fourier Analysis and related techniques to investigate the question of periodicities in fossil biodiversity. These techniques are able to identify cycles superimposed on the long-term trends of the Phanerozoic. We review prior results and analyze data previously reduced and published. Joint time series analysis of various reductions of the Sepkoski Data, Paleobiology Database, and Fossil Record 2 indicate the same periodicity in biodiversity of marine animals at 62 Myr. We have not found this periodicity in the terrestrial fossil record. We have found that the signal strength decreases with time because of the accumulation of apparently "resistant" long-lived genera. The existence of a 62 Myr periodicity despite very different treatment of systematic error, particularly sampling-strength biases, in all three major databases strongly argues for its reality in the fossil record (Melott and Bambach 2009a).

Part 2:
We use Fourier Analysis to investigate geological and isotopic data possibly related to periodicities in fossil biodiversity. Testing for cycles similar to the 62 Myr cycle in fossil biodiversity superimposed on the long-term trends of the Phanerozoic as described in Part I, we find a significant (but weaker) signal in sedimentary rock packages, particularly carbonates, which suggests a connection. Coincidence in timing is more consistent with a common cause than a basis in sampling bias. We find that Exxon sea level shows no significant periodicity, but one component of its fluctuation is consistent in period and phase with the biodiversity and sedimentation periodicities. A previously identified set of mass extinctions are found to lie preferentially on the declining phase of the 62

Myr periodicity, supporting the idea that the periodicity relates to variation in biotically important stresses. When detrended, the ratio of $^{87}Sr/^{86}Sr$ in sea water shows a strong periodicity near the same period, but almost perfectly out of phase: high values are nearly coincident with the low values of the periodicity in biodiversity. Periodic increase in continental freeboard could account for all these ~60 Myr periodic phenomena. Further work should focus on finding the underlying cause of the 62 Myr periodicity that links fossil biodiversity, sedimentary packages, and strontium isotope ratios (Melott and Bambach 2009b).